# Magnetic properties of Cobalt films described by second order perturbed Heisenberg Hamiltonian

P. Samarasekara and Amila D. Ariyaratne

Department of Physics, University of Peradeniya, Peradeniya, Sri Lanka

## Abstract

Second order perturbed Heisenberg Hamiltonian was employed to investigate the magnetic properties of hexagonal Cobalt films. Initially the number of nearest neighbors and the constants arisen from the partial summation of the dipole interactions of the structure of cobalt were calculated using some special algorithms. Minimization of the energy difference between the easy and hard direction of a memory device is very important. When the energy difference between the easy and hard directions is significantly small, the magnetic moments in a memory device can be quickly rotated between easy and hard directions under the influence of a small magnetic field. The thickness of a cobalt film corresponding to this minimum energy difference calculated using this theoretical model agrees with some experimental data of cobalt based magnetic memory devices.

## I. INTRODUCTION

Thin films of cobalt based materials find potential applications in memory discs and storage devices [1]. But any theoretical investigation related to the ferromagnetic cobalt thin films have not been performed by any researcher. The magnetic properties of ferromagnetic thin and thick films with simple cubic (sc), body centered cubic (bcc) and face centered cubic (fcc) have been explained using oriented, second order perturbed and third order perturbed Heisenberg Hamiltonian by us previously [2-4, 8]. The easy and hard directions have been determined in each case. According to those studies, easy and hard direction of ferromagnetic thin films depend on magnetic exchange interaction, dipole interaction, second and fourth order anisotropy, demagnetization factor, magnetic field and stress induced anisotropy. In all above cases, c-axis of the lattice was assumed to be perpendicular to the substrate. In addition, the ferromagnetic properties of Fe and Ni have been explained using a similar model by some other researchers [5, 6].

Due to the complexness of the hexagonal closed packed (hcp) structure, the determination of constants arisen from the partial summations of dipole interactions were complicated compared to determination of those of sc, fcc and bcc lattices. So an algorithm has been employed to evaluate these constants. This same strategy has been applied to determine these constants of Nickel ferrite by us previously [7, 11, 12, 13, 14]. According to experimental data, the stress induced anisotropy of ferrite films is considerable [9, 10, 15].

**II. THE MODEL**

The Heisenberg Hamiltonian of any ferromagnetic film can be basically expressed as following.

$$H = -\frac{J}{2}\sum_{m,n}\vec{S}_m \cdot \vec{S}_n + \frac{\omega}{2}\sum_{m \neq n}\left(\frac{\vec{S}_m \cdot \vec{S}_n}{r_{mn}^3} - \frac{3(\vec{S}_m \cdot \vec{r}_{mn})(\vec{r}_{mn} \cdot \vec{S}_n)}{r_{mn}^5}\right) - \sum_m D_{\lambda_m}^{(2)}(S_m^z)^2 - \sum_m D_{\lambda_m}^{(4)}(S_m^z)^4$$

$$-\sum_{m,n}[\vec{H} - (N_d \vec{S}_n / \mu_0)] \cdot \vec{S}_m - \sum_m K_s \sin 2\theta_m \qquad (1)$$

For a thick ferromagnetic film, the solution of above equation can be given as,

$$E(\theta) = -\frac{J}{2}[NZ_0 + 2(N-1)Z_1] + \{N\Phi_0 + 2(N-1)\Phi_1\}\left(\frac{\omega}{8} + \frac{3\omega}{8}\cos 2\theta\right)$$

$$- N(\cos^2\theta D_m^{(2)} + \cos^4\theta D_m^{(4)} + H_{in}\sin\theta + H_{out}\cos\theta - \frac{N_d}{\mu_0} + K_s \sin 2\theta)$$

$$-\frac{[-\frac{3\omega}{4}(\Phi_0 + 2\Phi_1) + D_m^{(2)} + 2D_m^{(4)}\cos^2\theta]^2 (N-2)\sin^2 2\theta}{2C_{22}}$$

$$-\frac{1}{C_{11}}[-\frac{3\omega}{4}(\Phi_0 + \Phi_1) + D_m^{(2)} + 2D_m^{(4)}\cos^2\theta]^2 \sin^2 2\theta \qquad (2)$$

In above equation; N, J, $Z_{|m-n|}$, $\Phi_{|m-n|}$, $\omega$, $\theta$, $D_m^{(2)}$, $D_m^{(4)}$, $H_{in}$, $H_{out}$, $N_d$, $K_s$ are total number of layers, spin exchange interaction, number of nearest spin neighbors, constants arisen from partial summation of dipole interaction, strength of long range dipole interaction, azimuthal angles of spins, second order anisotropy, fourth order anisotropy, in plane applied field, out of plane applied field, demagnetization factor and the stress induced anisotropy factor, respectively.

Here $C_{11}$ and $C_{22}$ are given by,

$$C_{11} = JZ_1 - \frac{\omega}{4}\Phi_1(1+3\cos 2\theta) - 2(\sin^2\theta - \cos^2\theta)D_m^{(2)}$$
$$+ 4\cos^2\theta(\cos^2\theta - 3\sin^2\theta)D_m^{(4)} + H_{in}\sin\theta + H_{out}\cos\theta - \frac{2N_d}{\mu_0} + 4K_s\sin 2\theta$$

$$C_{22} = 2JZ_1 - \frac{\omega}{2}\Phi_1(1+3\cos 2\theta) - 2(\sin^2\theta - \cos^2\theta)D_m^{(2)}$$
$$+ 4\cos^2\theta(\cos^2\theta - 3\sin^2\theta)D_m^{(4)} + H_{in}\sin\theta + H_{out}\cos\theta - \frac{2N_d}{\mu_0} + 4K_s\sin 2\theta$$

### III. RESULTS AND DISCUSSION

The diagram of conventional unit cell of cobalt with lattice parameters (a and c) is given in figure 1. The c/a ratio for Cobalt is 1.62.

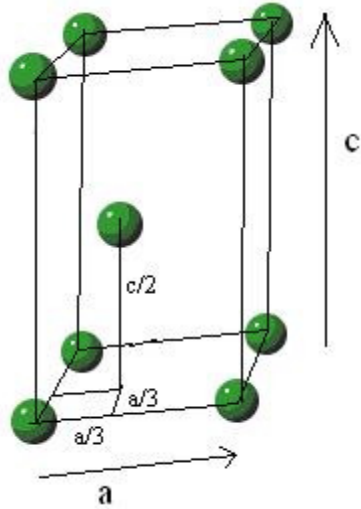

**Fig. 1:** Conventional unit cell of cobalt

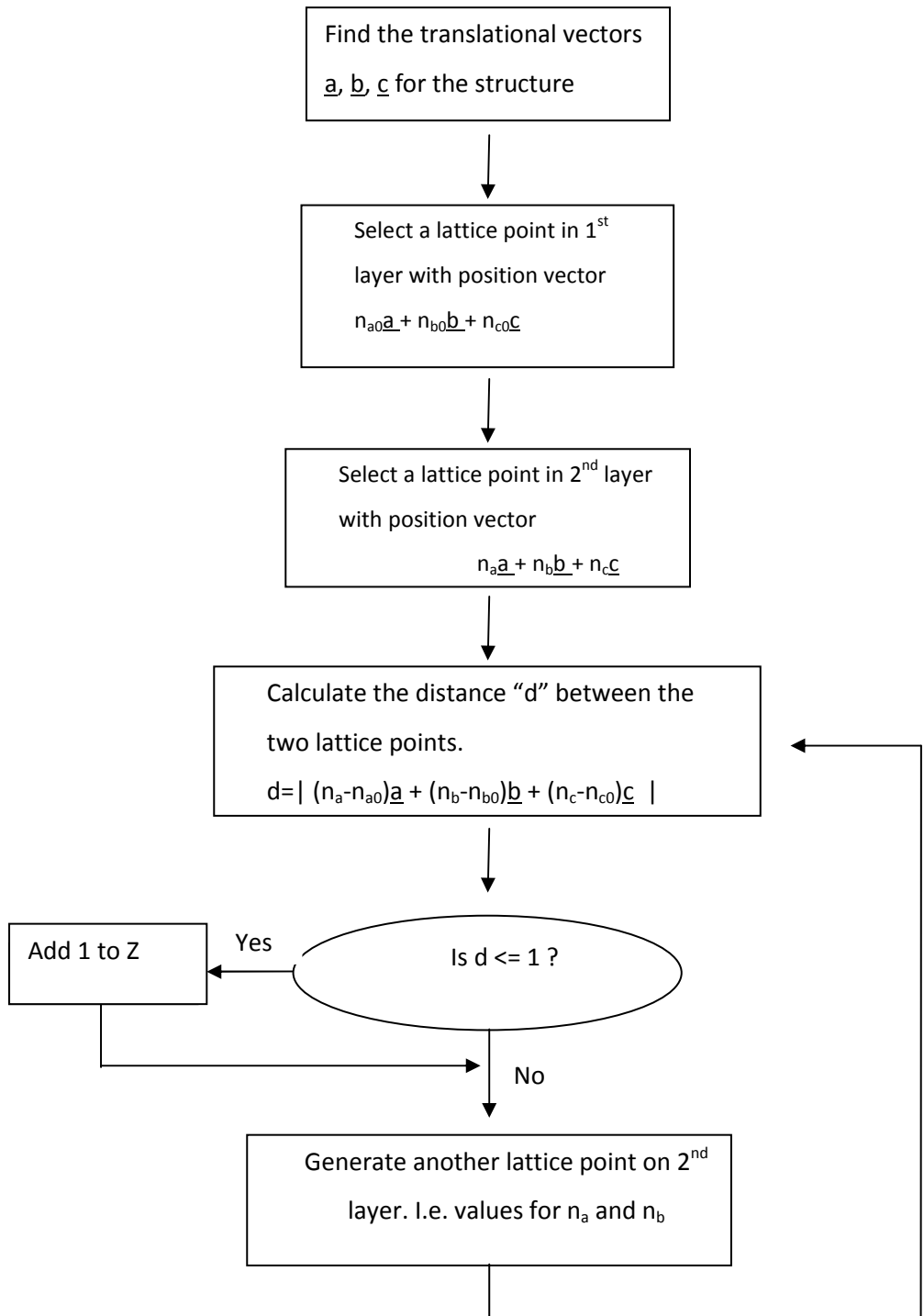

**Fig. 2:** Algorithm to calculate Z

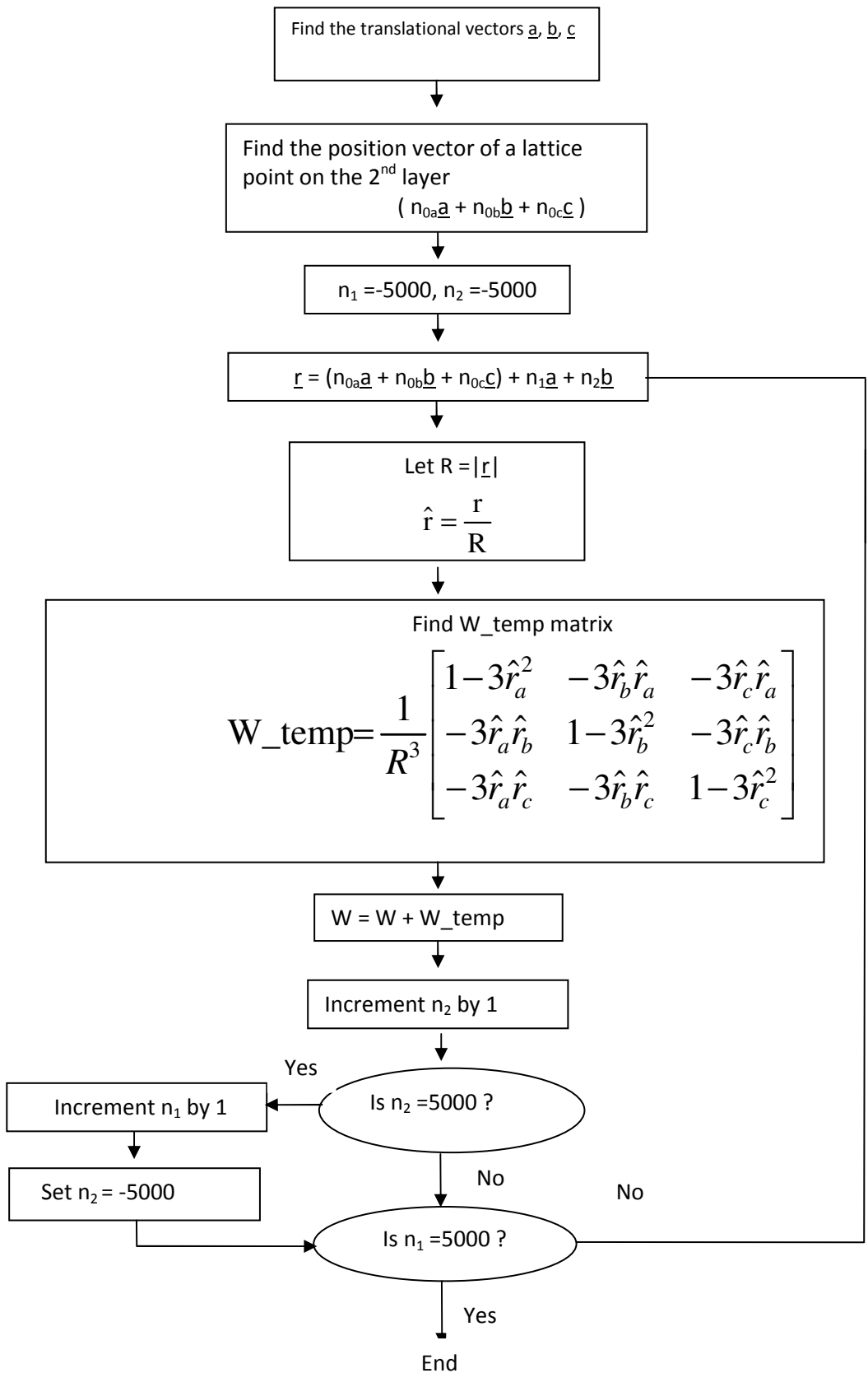

**Fig. 3:** Algorithm to calculate Φ

Then the algorithm given in figure 2 was implemented to evaluate the number of nearest neighbors in a cobalt film with spin layers parallel to the substrate. The 2nd algorithm given in figure 3 was applied to calculate the constants arisen from the partial summations of dipole interactions. Finally following values were found for hcp lattice.

The number of nearest neighbors in one lattice plane=$Z_0$=6

Number of nearest neighbors between two adjacent lattice planes=$Z_1$=3

Constants arisen due to the partial summation of dipole interactions in one layer=$\Phi_0$=11.0324

Constants arisen due to the partial summation of dipole interactions between two adjacent layers=$\Phi_1$=0.4210

Since the experimental values of $D_m^{(2)}, D_m^{(4)}, H_{in}, H_{out}, K_s$, J and ω have not been measured for cobalt thin films by any researcher yet, the simulations were carried out for a reasonable set of $D_m^{(2)}, D_m^{(4)}, H_{in}, H_{out}, K_s$, J and ω values as given below.

$$\frac{J}{\omega} = \frac{D_m^{(2)}}{\omega} = \frac{H_{in}}{\omega} = \frac{H_{out}}{\omega} = \frac{N_d}{\mu_0 \omega} = \frac{K_s}{\omega} = 10 \quad and \quad \frac{D_m^{(4)}}{\omega} = 5$$

The graph of $\frac{E(\theta)}{\omega}$ versus angle is plotted in Figure 4. As indicated, the easy direction is found at an angle of about 40° and the hard direction occurs at an angle of about 140°. For a material with a simple structure, the angle between the easy and hard directions is $90^0$. But the angle between easy and hard directions is $100^0$ in this case due to the complexness of the structure of cobalt.

The Figure 5 shows the variation of the energy difference between the easy direction and the hard direction against the number of layers for a Cobalt thin film. It could be observed from this graph that the energy difference is a minimum for a film of 50 layers. Therefore, a hard disk drive would require a less amount of energy to store data if the magnetic film is synthesized with the number of layers being in the above region. This theoretical result agrees with that of modern hard disks. The optimum experimental results for Co based magnetic memory devices have been obtained for a thin film with the same number of layers by some other researchers [1].

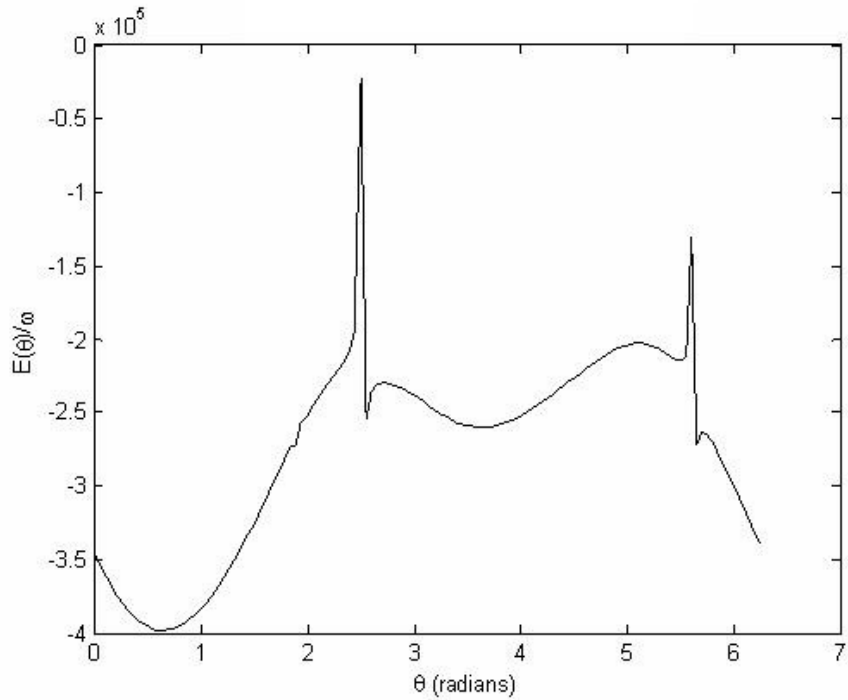

**Fig. 4:** The graph of $\dfrac{E(\theta)}{\omega}$ versus θ for a Cobalt thin film with 5000 layers

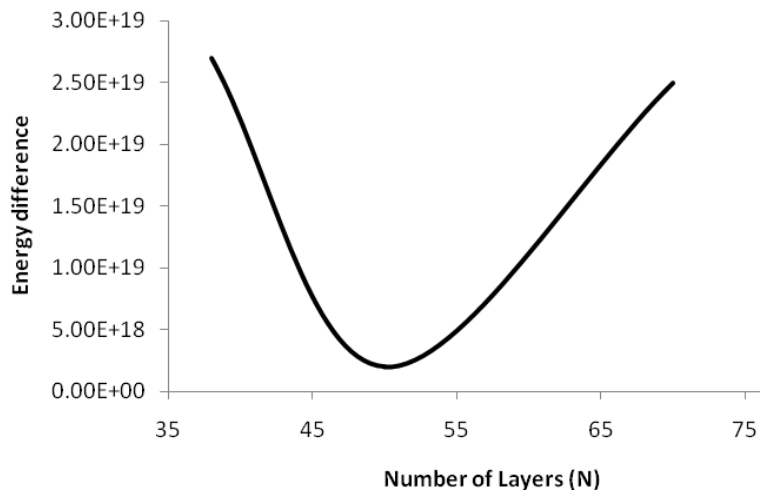

**Fig. 5:** The variation of the energy difference between the easy and the hard directions against the number of layers N

3-D plot of $\frac{E(\theta)}{\omega}$ versus angle and number of layers is given in figure 6. The angle and number of layers corresponding to easy and hard directions can be determined using this plot. The difference between the maximum and minimum energies is really small around N=50 according to this graph too. So the energy required to rotate from easy to hard direction (crystal anisotropy) is significantly small for films with 50 layers.

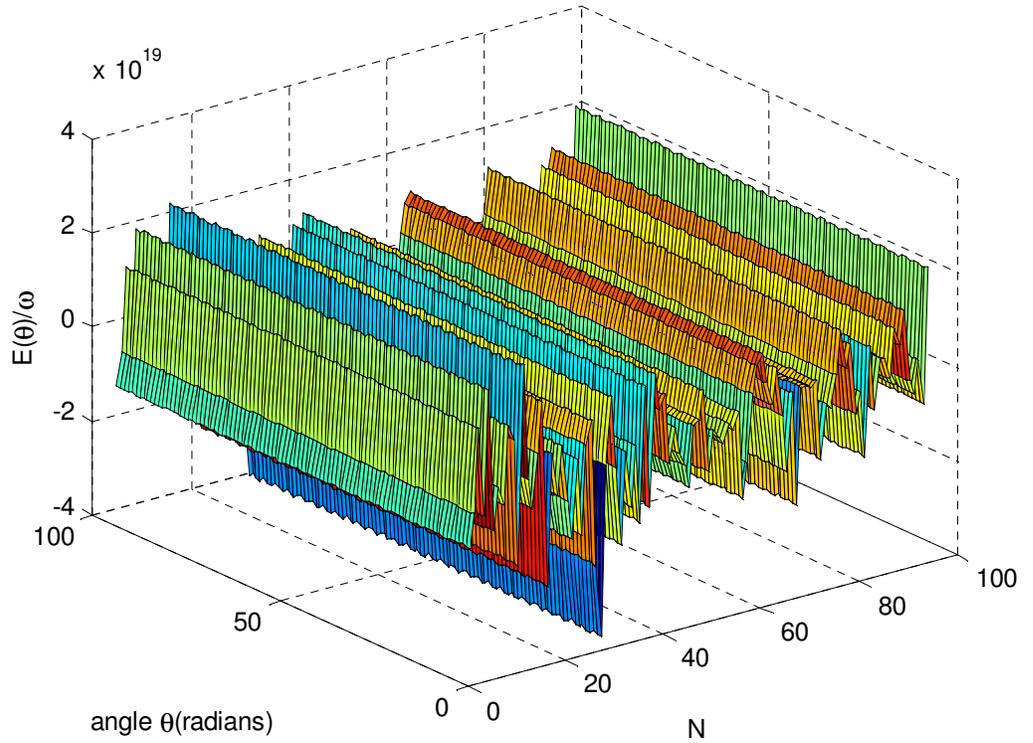

**Fig. 6:** 3-D plot of $\frac{E(\theta)}{\omega}$ versus angle (θ) and number of layers (N)

## IV. CONCLUSION

Values of number of nearest neighbors and the constants arisen from the partial summations of dipole interactions calculated using the algorithms given in figures 2 and 3 are $Z_0=6$, $Z_1=3$, $\Phi_0=11.0324$ and $=\Phi_1=0.4210$ for cobalt thin films. This simulation was carried out for a selected set of values of energy parameters in order to study the variation of total magnetic energy with angle (θ) and the number of layers (N). According to the energy curves, the energy difference between the easy and hard directions can be minimized at N=50. This number of layers (N=50) is approximately equal to the thickness of cobalt films synthesized for magnetic memory applications by some other researchers [1]. This implies that the magnetic moments of a cobalt based memory device can be easily rotated between easy and hard directions, when the number of layers is 50.